# The Hierarchical Potential Energy Landscape of Screw Dislocation Motion in Refractory High-entropy Alloys

Xinyi Wang[a,b], Francesco Maresca[c,*], Penghui Cao[a,b*]

[a]Department of Mechanical and Aerospace Engineering, University of California, Irvine, CA 92697, USA

[b]Materials and Manufacturing Technology Program, University of California, Irvine, CA 92697, USA

[c]Engineering and Technology Institute Groningen, Faculty of Science and Engineering, University of Groningen, 9747 AG, Netherlands

[*]To whom correspondence may be addressed.
Emails: f.maresca@rug.nl; caoph@uci.edu

This article contains supporting information.

## Abstract

High-entropy alloys (HEAs) with concentrated solid solutions are conceived to possess a rugged atomic and energy landscape in which dislocation motion necessarily proceeds to accommodate mechanical deformation. Fundamental questions remain as to how rough the energy landscape is and to what extent it can be influenced by the local ordering of their constituent elements. Here, we construct and report the potential energy landscape (PEL) governing screw dislocation motion in refractory HEAs that reveals a hierarchical and multilevel structure with a collection of small basins nested in large metabasin. This striking feature pertaining to HEAs exerts a trapping force and back stress on saddle point activations, retarding dislocation movement. By introducing chemical short-range order, the energy landscape is smoothed but skewed to different degrees that shifts the rate-liming process from kink-glide to kink-pair nucleation. The chemical disorder-roughened PEL in HEAs, analogous to structural disorder induced in metallic glasses, signifies the role of various barrier-hopping processes underlying the extraordinary mechanical behaviors of the two distinct groups of materials.

**Introduction**

    Dislocations, the line defects in a crystal lattice, dictate the strength and deformation behaviors of materials[1]. Their glide that must overcome an energy barrier due to lattice resistance can be facilitated by applied mechanical stress, thermal activation, or both of them. At zero temperature, a minimum stress (Peierls stress) is required to vanish the energy barrier (Peierls barrier) and move a straight dislocation in the absence of thermal activation[2,3]. As the temperature is raised, dislocation motion can occur via kink-pair mechanism at a stress level below the threshold value[4]. In pure metals, the three (constant) energy barriers associated with Peierls mechanism, kink-pair nucleation, and kink gliding render a smooth potential energy landscape (PEL) that regulates the dislocation movement. Because of the easy glide of edge dislocation in body-centered cubic (bcc) metals, the sluggish motion of screw resulting from the kink-pair process (considerably high energy barrier) is the essential origin of high strength[5]. The promotion of kink-pair nucleation and increased concertation of thermal kinks, with increasing temperature, would smear out the rate-limiting effects of the kink-pair mechanism, which is closely correlated to the apparent temperature dependence of strength of bcc metals[6,7]. In dilute bcc alloys, the addition of solute atoms that lowers the double kink nucleation barrier and increases the barrier for kink lateral migration[8,9] would roughen and tilt the PEL, in which kink-pair activation process is favored and kink propagation is impeded.

    In contrast to bcc pure metals and dilute alloys, the refractory high entropy alloys (HEAs)[10] such as MoNbTaW[11] and MoNbTi[12], exhibit a gradual and slow decrease in strength with increasing temperature to half of their melting points. The high strength and its weak temperature dependence are ascribed to dislocations that operate in concentrated solid solutions. The locally high chemical fluctuations in HEAs inevitably influence the element mechanisms of dislocation motion and their corresponding energy barriers, and hence the structure of PEL. It is conceivable that a high PEL ruggedness and large variation of energy barriers can be induced by the composition fluctuations, and impact the dominance of individual mechanisms governing dislocation motion. For example, the wide distribution of dislocation core energy indicating the high variation of Peierls valley level reflects the rugged energy landscape in HEAs[13]. Indeed, the local chemical fluctuations and their interaction with the dislocation can be so strong to give rise to changes of the local dislocation core structure[14]. Since the dislocation energy merely presents

local minima in the PEL, important questions remain as to how rough the energy landscape is and to what extent the energy barrier varies in HEAs.

Another salient and emergent feature pertaining to HEAs that makes them distinct from traditional alloys is chemical short-range order (SRO)[15]. The actual materials, which are processed, homogenized or annealed at temperatures below their melting points, unavoidably accommodate SRO, because of the attractive/repulsive interactions among the constituent elements. Especially when the mixing enthalpy becomes the predominant term in the Gibbs free energy, the diffusion kinetics leading to a lower free-energy state renders the formation of SRO inevitable[16,17]. It has been shown that, in fcc alloys, SRO can have a considerable impact on dislocation motion and modifies its structure from wavy to planar[18], impacting the mechanical behaviors, e.g., slip resistance[19] and hardness[16]. In the context of bcc HEAs, the role of SRO on individual mechanisms underpinning dislocation motion and the PEL structure has yet to be elucidated.

This study samples and reconstructs the energy landscape formed by concentrated solid solutions and reveals the saddle point activation events along minimum energy pathways that trigger individual mechanisms (Peierls mechanism, kink-pair formation, and lateral kink glide) underlying screw dislocation motion. Considering a model bcc MoNbTaW alloy using a state-of-the-art machine learning interatomic potential[20] that captures solute interactions and ordering, the potential energy landscape demonstrates hierarchical structures with large metabasin encompassing a bundle of small local basins - a striking feature resulting from random HEA solid solutions. The energy landscape, sensitive to local chemical order, can be altered by introducing SRO, which transfers the predominance of individual mechanisms as the rate-limiting step for screw dislocation motion. The detailed analysis of the PEL in the random solid solution (RSS) alloy shows that, in contrast with both the very recent understanding[21,22] of screw dislocation glide, kink-pair nucleation can be a rate-limiting mechanism that competes with kink pair propagation in concentrated solid solutions. The PEL of the SRO system highlights the various roles of breaking chemical order and consequent anti-phase boundary (APB) generation in modifying the energy barriers of elementary gliding processes.

## Methods

**MC/MD simulations and chemical SRO parameter.** We use a Monte Carlo (MC) swap of atoms coupled with MD simulations to prepare the NbMoTaW with lowered potential energy and increased SRO. The MC/MD simulation[23] is performed at 300 K with Nose-Hoover thermostat and involves 1,000,000 MD steps and calls MC trials every 1,000 timesteps to perform trials of exchanging each pair of elements. The atom swaps are accepted or rejected based on the Metropolis algorithm. To measure chemical SRO, the non-proportional number is used to quantify the degree of chemical order[24]. The order parameter between any pair of atoms $i$ and $j$ is defined as $\delta_{ij}^k = N_{ij}^k - N_{0,ij}^k$, where $N_{ij}^k$ denotes the actual number of pairs in $k$th shell, and $N_{0,ij}^k$ the number of pairs for the pure random mixture. Following this definition, we can see a positive $\delta_{ij}^k$ indicates a favored and increased number of pairs, meaning element $i$ tends to bond with element $j$ in the $k$th shell, while a negative value represents an unfavored pairing. In the Supplementary Figure S1, we present the MC/MD simulation results, atomic configurations, and the calculated order parameters $\delta_{ij}$ for the first nearest neighboring shell. The order parameter values of all individual pairs in RSS system are zero, indicating the random nature. The MC/MD annealed structure exhibits strong ordering between Mo-Ta ($\delta_{Ta-Mo}^{k=1} = 2.044$), Nb-Mo ($\delta_{Nb-Mo}^{k=1} = 1.468$), and W-Ta ($\delta_{W-Ta}^{k=1} = 1.416$) pairs.

**MD simulations of screw dislocation motion in RSS and SRO.** To reveal screw dislocation motion behaviors in RSS and SRO, we perform MD simulations of a periodic array of dislocations subjected to constant shear stress. The crystal has dimensions of ~250 Å in $x$ ([11$\bar{2}$]), 240 Å in $y$ ([$\bar{1}$10]) and 168 Å in $z$ ([111]) directions, containing 580000 atoms. As shown in Supplementary Fig. S2, a screw dislocation with a length of $60b$ is introduced to the system along $z$ by imposing a linear displacement on all atoms in the upper half of the cell. The system orientations give rise to the glide direction is along $x$ and the glide plane normal direction $y$. Periodic boundary conditions are imposed along $x$ and $z$, and zero tractions on the two free surfaces parallel to the glide plane. To drive the motion of screw dislocation, we apply the constant shear stress to the system by adding a constant force to two atomic layers of upper and lower surfaces. A machine learning potential[20] is adopted to model interatomic interactions in NbMoTaW, which has been verified for predicting the non-degenerate core structure of screw dislocation.

**Peierls mechanism and kink-pair nucleation models.** We use the free-end nudged elastic band (NEB) method[25] to construct the minimum energy pathway of dislocation glide over two neighboring Peierls valleys. By moving the dislocation to next valley and repeating the same procedure, the energy landscape of dislocation glide over a large distance is obtained. For Peierls mechanism, a short dislocation of $4b$ length is considered and inserted in the system (see Supplementary Fig. S4). We minimize and connect the energy pathways spanning 25 Peierls valleys using NEB calculation. The NEB inter-replica spring constant is set to be $k = 0.01$ eV/A$^2$ and force tolerance 0.001 eV/A. The choice of this value that optimizes convergence of the calculations results in essentially the same energy barrier using smaller tolerance and large spring constant (Supplementary Fig. S6). In order to explore kink pair nucleation energy barriers, a longer dislocation with $20b$ is adopted, which naturally leads to one kink-pair event along the dislocation line. The kink-pair nucleation mechanism and the minimum energy pathways are calculated using NEB by considering 30 different sites in RSS and SRO, respectively.

**Kink glide model.** To construct dislocation with one single kink, we first build a crystal with orientations of $[\bar{9}\,9\,\overline{20}]$, $[110]$, and $[\overline{10}\,10\,\bar{9}]$ along $x$, $y$, and $z$ directions. A screw dislocation is inserted into the system with a line direction of $z$. The angle between $[\overline{10}\,10\,\bar{9}]$ and $[\bar{1}\,1\,\bar{1}]$ directions is 2.79°, and the distance between two Peierls valley is $0.94b$ ($b$ is Burgers vector). The dislocation length for forming one single kink is 54.25 Å in the $z$ direction (see Supplementary Figure S5). After energy minimization, one single kink is formed in the relaxed and kinked dislocation line locating in the Peierls valleys (i.e., along with the $[\bar{1}\,1\,\bar{1}]$ direction). We then perform NEB calculation and construct the minimum energy pathway of kink-glide connecting two neighboring energy minima. By sampling a large glide distance of $650b$, the potential energy landscape governing kink propagation is obtained in RSS. As to SRO, the initial crystal is annealed through MC/MD simulation, followed by the same procedure, i.e., $[\overline{10}\,10\,\bar{9}]$ orientated dislocation generation and minimum energy path calculation.

## Results

**Phenomenology of screw dislocation motion in RSS and SRO.** We first report screw dislocation gliding behaviors in the two different chemical environments, i.e., RSS and SRO. The length of the modeled dislocation is $60b$ ($b$ is the Burgers vector magnitude), which is large enough to enable the three elemental mechanisms (Supplementary Fig. S2). Figure 1a-b shows the typical sequences of a screw dislocation gliding in RSS and SRO, respectively, at T=300 K and 800 MPa applied shear stress. The evolution of dislocation morphologies shows that the dislocation line in RSS is more rugged, and many kinks persist as the motion progresses. Since the screw dislocation line vector and the Burgers vector are parallel, a glide plane is not defined, and the screw dislocation can move in any of the three {110} planes that contain the Burgers vector. In Figure 1a-b, we use various colors to indicate dislocation slip on different planes.

By examining the detailed dislocation motion (Figure 1a), two kink-pairs are nucleated within the first 4 ps from the initially straight dislocation, moving part of the dislocation line two Peierls valleys away and on a different plane than the maximum resolved shear stress (MRSS) plane. Lateral kink migration of the two kink-pairs stops at 7.5 ps, suggesting that a large energy barrier has been encountered. The further advancement of the dislocation proceeds with nucleation of a third kink-pair in another region. The new kink pair expands rapidly and one of its kinks recombines with the immobile kink, overcoming the local barrier which was retarding the dislocation motion. At 10.5 ps, the entire dislocation has moved away from its initial Peierls valley after further lateral kink glide and annihilation of opposite-sign kinks. At 11.25 ps, three more kink-pairs nucleates, one of which is on a different (red) plane. Two kinks belonging to different planes collide and generate a cross-kink, leading to screw dislocation self-pinning (12 ps). The dislocation unpins from cross-kinks requiring a long-waiting time implies a strong self-pining effect of the cross-kink (Supplementary Fig. S3). In contrast to RSS, dislocation motion in the presence of SRO reveals an overall less rugged structure in the displaced configurations, involving fewer kink-pairs (Figure 1b). At 5.5 ps, a kink-pair forms on the MRSS plane, and the kinks migrate along the dislocation line (7.75 ps) to eventually recombine, as one would expect to occur in an elemental bcc metal. After the entire dislocation has moved to the next Peierls valley (10 ps), the same processes (kink-pair nucleation and kink propagation) repeats, yet involving kink-pair nucleation on a different plane (18.75 ps). Cross-kink formation is not observed for this $60b$ long dislocation after simulating glide for 500 ps.

The collision of kink-pairs on different slip planes results in cross-kinks that act as self-pinning points. Depending on the applied stress and temperature, cross-kinks can be overcome by two distinct mechanisms: (1) stress-driven cross-kink unpinning and (2) thermally-activated and diffusion-mediated cross-kink annihilation. Figure 1c elucidates the atomistic processes underlying cross-kink formation and breaking at low/moderate temperatures (300 K). The unpinning process resulting from stress-induced cross-kink breaking produces debris that is left behind the moving dislocation and that consists of vacancy and self-interstitial loops[20]. At high temperatures (900 K), the second type of mechanism occurs, as shown in Figure 1d. The two cross-kinks diffuse along the dislocation line (the interstitial-type kink drifts faster) and interact with each other. Eventually, the (cross-kink) vacancy and self-interstitial recombine, thus annihilating the cross-kinks. This thermally-activated disentanglement, which is enabled by the high-temperature diffusion, suggests that the self-pinning strengthening might become weakened with increasing temperature, calling for other strengthening mechanisms sustaining the strength at high temperatures[27].

The screw dislocation motion in RSS and SRO environments reveals that distinct mechanisms arise at different length- and time-scales. The spatial and temporal scales of dislocation motion mechanisms, regulated by their energy barriers and the PEL structure, are smaller and more frequent in the RSS case, possibly originating from the faster chemical and energy fluctuations compared with SRO. Next, we focus on the individual processes, i.e., the Peierls mechanism, kink-pair nucleation, and kink glide that mediate dislocation motion and reveal their underlying energy landscapes in SRO and RSS, with an attempt to isolate respective roles of concentrated solid solutions and local chemical order.

**Peierls mechanism and its energy landscape**. The Peierls mechanism, involving glide of a straight dislocation segment from one Peierls valley to the neighboring one, occurs via Peierls hill (barrier) crossing on PEL. To reveal its energy landscape and how SRO alters it, we consider a short screw dislocation with $4b$ length. For such a short-length segment, kink-pair nucleation is prevented, and the entire dislocation can glide over the Peierls. Figure 2a shows the distance traveled by the screw dislocation as a function of time (glide distance-time curves) in both RSS and SRO cases, subject to various shear stresses at 300 K. Irrespective of the applied stress, the dislocation glides faster in RSS than SRO. By tracking the dislocation core position upon gliding (Figure 2b), at the same stress level, the dislocation in RSS experiences more cross slip events than SRO. This appears from the larger vertical displacement component ($Y$-axis) of the dislocation core, hence a larger slope in the $X - Y$ plane (0.31 for RSS and 0.10 for SRO). These two findings suggest that the presence of SRO retards the Peierls mechanism and reduces the cross-slip frequency.

We now investigate the mechanistic origin of the reduced dislocation mobility by analyzing PEL and the Peierls barrier using minimum energy path calculation (Methods and Supplementary

Fig. S4). Figure 2c shows the constructed minimum energy pathway spanning 25 Peierls valleys for both SRO and RSS configurations. The potential energy (per Burgers vector) variation is plotted along with the dislocation distance over $25a$ (Peierls valleys spacing $a = 0.94b$). It can be seen that the hieratical structure emerges from the energy landscape with a series of local basins (Peierls valleys) embedding in large metabasins, a feature resulting from concentrated random solutions (upper panel of Figure 2c). Remarkably, the SRO energy landscape shows an increasing trend as the dislocation advances (middle panel of Figure 2c). This trend reflects that the chemical order on the slip plane is destroyed when the screw dislocation glides through the SRO crystal. The breaking of SRO, also known as APB generation, incurs an energy cost, thus resulting in an overall energy increase. By subtracting the APB energy (Supplementary Section 1), we obtain the energy landscape of the Peierls mechanism associated with the chemistry fluctuations (green line in Figure 2c). While the energy pathway for SRO after subtracting the APB energy shows a similar trend as in the RSS case, the dislocation core energies (local minima) have a more concentrated distribution in SRO than RSS (Supplementary Section 2). The SRO shrinking core energy (reduced variation of Peierls valley) is consistent with the recent study using density functional theory calculations[13].

Figure 2d shows the energy barrier distributions in the RSS, SRO, and SRO-APB (SRO after removing APB energy). The SRO shows an overall higher energy barrier than RSS, implying an enhanced lattice resistance to dislocation glide in a chemically ordered environment. After removing the APB energy contribution (bottom panel of Figure 2d), the barriers exhibit a similar distribution as in RSS: the mean value is similar, but the standard deviation is smaller (being 28.2 meV/b for SRO-APB and 41.8 meV/b for RSS case). This demonstrates the predominant role of breaking local chemical order and the generation of APB in increasing the Peierls barrier. It is worth noting that the average Peierls barrier in RSS and SRO-APB (103 meV/b) is close to the barrier estimated by averaging the elemental contributions (96 meV/b, see Supplementary Fig. S7), while SRO raises the value to 127 meV/b.

The simulation results and energy pathway analysis signify that the local chemical order introduced into refractory HEAs impacts the Peierls mechanism. The mobility reduction is associated with an increase in Peierls energy barrier connecting two energy valleys. As the dislocation jumps to the next valley, localized slip takes place and interrupts the chemical ordering on the slipped area, incurring an extra energy cost and hence raising the energy barrier. Concerning cross slip, the presence of SRO suppresses the frequency of cross slip and restrains the dislocation glide to one primary slip plane. The less frequent occurrence of cross-slip in the presence of SRO can be related to the smaller solute-dislocation interaction energy fluctuations[21] and to the energy cost of creating APBs, which might be larger in the cross-slipped case. In the RSS case, which encompasses more significant local chemistry (energy) fluctuations, the chance of finding an energetically more favorable environment on a different slip plane is also enhanced (originating from a wide distribution of dislocation core energy). Therefore, once the dislocation segment encounters on the MRSS plane a strongly unfavorable chemical environment that is hard to glide through, it prefers to change the glide plane towards a more favorable chemical environment. In contrast, the local environments in SRO are more uniform, and it is less likely to find a much lower energy environment on a slip plane different than the MRSS plane.

**Kink-pair nucleation and its energy landscape**. The kink-pair nucleation overcoming a sizable (constant) energy barrier in bcc pure metals is the rate-limiting step for screw dislocation motion. To inspect how the high solute concentration and SRO influence the kink-pair mechanism, we choose a screw dislocation with $20\,b$ length, a size that is long enough to prevent Peierls mechanism but appropriate to enable just one single kink-pair nucleation event (Methods and Supplementary Fig. S4). Figure 3a-b illustrates the schematics of the kink-pair nucleation process in RSS and SRO cases, respectively. It is noted that the kink-pair nucleation and propagation in SRO would destroy the chemical order in the kinked region (dislocation swept area). This disruption of local order should be accounted for when considering kink-pair nucleation in SRO environment.

Turning to energy barrier calculations of kink-pair nucleation, we show the typical energy landscape of the dislocation gliding to a neighboring Peierls valley via the kink-pair mechanism (Figure 3c). The migration pathway consists of three stages: (region i: A→B) metastable kink-pair nucleation (saddle point), (region ii: B→D) kink gliding, and (region iii: D→E) kink-kink annihilation, which happens naturally in a periodic simulation cell and that generally occurs on a long dislocation line, where multiple kinks are created, and opposite-signed ones can annihilate. A series of snapshots of the atomic configuration evolution (kink-pair and screw motion) along the pathway is presented in the right panel of Fig. 3c. Starting from the initial Peierls valley (energy minimum, A), a short line segment progressively glides to the next Peierls valley; at state (B), the kink-pair nucleation is complete and results in a metastable configuration. Thereafter, (C) the two kinks glide in opposite directions along the dislocation line. The lateral migration proceeds until (D) the two kinks attract each other due to opposite signs and (E) eventually annihilate (relax to next local energy minimum), resulting in a straight dislocation segment lying in a new Peierls valley. Figure 3d shows the minimum energy pathways associated with kink-pair mechanism for 30 independent calculations. We extract the energy barriers for critical-sized double-kink configurations in RSS and SRO, and show their distributions in Figure 3e. The SRO leads to an overall larger barrier than RSS. Thus, the kink-pair mechanism is more energetically costly in the presence of SRO than in RSS. This finding agrees with our direct long dislocation simulations, in which the screw dislocation motion in SRO exhibits fewer kinks (Figure 1b).

From the elastic interaction model[28], the excess energy associated with kink-pair nucleation in an elemental bcc crystal is $W_{KP} = 2E_{kink} + E_{int}$, where $E_{kink}$ is the (positive) kink self-energy and $E_{int}$ is the (negative, hence attractive) elastic interaction of the two opposite-signed kinks. In the case of an alloy, the bowed-out screw dislocation segment will generally experience a solute-dislocation interaction (or segregation) energy[21,21], and a new term $\Delta E_{seg}$ that accounts for the changes in local energy when a segment travels to the next Peierls valley should be considered. Therefore,

$$W_{KP}^{RSS} = 2E_{kink} + E_{int} + \Delta E_{seg}$$

can be conceived as a reasonable approximation of the RSS energy cost due to kink-pair formation (depicted in Figure 3a). In the presence of SRO, the APB contribution should also be accounted for the area swept by the kink pair (Figure 3b), hence

$$W_{KP}^{SRO} = 2E_{kink} + E_{int} + \Delta E_{seg} + E_{apb}$$

and the APB contribution (disruption of local order) $E_{apb}$ is always positive. Due to the local variation in solutes-dislocation interactions, it is expected that, on a long dislocation line, the nucleation site should be energetically favorable for segment advancement and hence $\Delta E_{seg}$ is likely negative, in both RSS and SRO cases. However, in the SRO case, it is apparent from the data shown in Figure 3 and the analysis above that segment advancement during kink-pair mechanism locally breaks the chemical order and incurs an extra energy cost because of the APB formation (area swept by the segment advancement). In the bottom panel of Figure 3c, we present the kink-pair formation energy distribution after subtracting this APB energy from the SRO data. Interestingly, the reduced distribution still shows a slightly higher value but a more compact distribution compared with RSS. Since the average critical kink-kink separation is essentially unchanged between RSS and SRO (~5.9$b$ in both cases), a plausible explanation for the difference is a larger kink-pair formation energy fluctuations in RSS compared with SRO, reflected by a larger standard deviation in the kink-kink separation for RSS (1.1$b$) compared with SRO (0.8$b$) (a rough energy landscape encountered by a kink is confirmed in the next Section). Another discrepancy between RSS and SRO-APB can originate from the typical energy changes $\Delta E_{\text{seg}}$ experienced after segment hopping. In the RSS system, the screw dislocation is more likely to find a more favorable kink-pair nucleation site that is associated with a larger solute-dislocation interaction-energy decrease due to strong chemical fluctuations.

**Kink glide and its energy landscape**. In pure bcc metals, lateral kink migration along the screw dislocation line has a periodicity of $b$ (magnitude of Burgers vector), and the activation energy for lateral glide is negligible so that kinks can glide easily, without thermal activation. The highly concentrated solid solute in HEAs is expected to appreciably change the kink-glide. Here, we construct a screw dislocation containing a single kink and carefully inspect the energy landscape over which the kink must travel (Methods and Supplementary Fig. S5). Figure 4b shows the computed minimum energy pathways of a single kink glide spanning $615b$ glide distance in both RSS and SRO. Along the pathway, the intermittent kink glide events are interspersed with kink-pair formation events. When the kink is trapped by favorable local environments or faces unfavorable environments ahead of its propagation direction, new kink-pair nucleates somewhere along the dislocation line and the screw dislocation can continue to glide (illustrated in Figure 4a).

In RSS (Figure 4b, top panel), the minimum energy pathway and landscape consist of a series of energy minima and saddle points. The landscape is highly rugged and appears hierarchical, featuring small basins and larger metabasins. Along the path, the two activation events, i.e., kink glide and kink-pair-nucleation, are denoted by filled and unfilled symbols, respectively. The migration of the existing kink will be the active mechanism (Figure 4c) if its glide barrier is smaller than that of kink-pair nucleation in the dislocation. Once the kink experiences a large energy barrier to glide over (induced by its local environments), the dislocation emits a pair of kinks, and their propagation resumes the movement of the dislocation towards the next Peierls. As one of the newly formed kinks propagates to the pinned kink, eventually, the kinks annihilate with the aid of the elastic kink-kink attraction (configuration evolution and pathway shown in Figure 4d). Thus, the moving screw dislocation in RSS exhibits both active kink glide and kink-pair nucleation. When the existing kink is trapped and pinned, kink-pair nucleation takes over and produces mobile kinks that resume dislocation motion.

The energy pathway of kink glide in SRO shows an increasing trend (the middle panel of Figure 4b). As kink glides, the forward glide of the screw dislocation breaks the chemical order on the swept area, incurring an extra energy cost. Intriguingly, the fraction of kink-pair events during kink glide is markedly reduced in the presence of SRO, implying enhanced kink glide and hindered kink-pair mechanism. Since SRO increases the kink-pair energy barrier over glide barrier, it would raise the importance of kink-pair nucleation as the rate-limiting process. After removing the energy increase associated with APB creation (breaking SRO) along the migration path, we obtain the energy landscape of the kink glide along with the screw dislocation. The PEL shows a longer length scale (periodicity) between local energy minima of kinked configuration compared to RSS (14.7b for SRO, 11.6b for RSS, see Supplementary Section 3), and kink nucleation is less frequent in the SRO environment (large formation energy shown in Figure 3e). We can then conclude SRO suppresses kink nucleation rate and favors kink glide as the mechanism facilitating dislocation motion.

**Discussion and conclusion**

The energy landscape structure that dictates dislocation motion in HEAs, revealed here by means of a minimum energy pathway sampling and a machine learning potential, exhibits a distinct structure from that of pure elements and traditional dilute alloys. Specifically, the trajectories of minimum energy migration pathway (Fig. 2c and Fig. 4b) suggest the HEAs possess a hierarchical energy landscape consisting of metabasins breaking into a collection of small sub-basins. The saddle point activations and elementary barrier hopping on the energy landscape, appearing as local processes such as dislocation segment advancement, kink-pair nucleation, and kink glide, regulate dislocation motion. When the system attempts to escape from one metabasin to an adjacent one, it goes through a series of local barrier activations and inevitably bears accelerated backward jumps during metabasin climbing (forward barrier larger than backward one). This emergent PEL feature, stemming from the multi-principal elements in high concentration, exerts a trapping force and back stress on saddle point activations, retarding dislocation movement. We regard the hierarchy of the energy landscape as the salient character of HEAs that makes them different from pure metals and dilute alloys. Interestingly, the hierarchical and fractal features are ubiquitous in glasses (the disordered materials that lack long-range structural order) and are considered to be the root cause of many anomalous behaviors of glassy material, for instance, rheological behavior[29]. In HEAs, however, the rugged PEL and its multilevel picture are induced by concentrated solid solutions and their chemical disorder. The resultant rugged PEL and its variability with chemical disorderliness act a vital role in dislocation motion and hence unusual strengthening in refractory HEAs.

When SRO is introduced into the system, the PEL of dislocation processes is tilted, because of the chemical order breaking and APB creation accompanied by individual dislocation advancement mechanisms. In the presence of SRO, the energy barriers for individual mechanisms (Perierls hopping, kink-pair, and kink-glide) all increase but to different extents. Because of various SRO sensitivities, kink-pair mechanism is appreciably impeded, yet kink glide is comparatively promoted, leading to the dominant mechanism drift towards kink-pair nucleation. This is signified regarding kink propagation and new kink-pair formation (Figure 4), where the SRO substantially reduces the occurrence of kink-pair nucleation (large energy barrier and slow process) and turns it into the predominated rate-limiting step for dislocation motion. The pure effects of SRO on dislocation core energy (Peierls valley), kink energy (local energy minimum), Peierls and kink gliding barriers (saddle points on PEL) are isolated by eliminating the energy penalty incurred by breaking local chemical order. The variances of local minimum and energy barrier distributions are shrunk, implying that SRO smoothens the PEL and reduces its ruggedness related to solute-dislocation interaction energy fluctuations.

Since the screw dislocation strength is the sum of the strengthening processes from segments/kinks gliding and cross-kinking[21,30], whether SRO will strengthen the alloy or not depends on specific compositions and their interaction with the dislocation. The presence of SRO brings about the back stress for local chemical order recovering but reduces variance in the distribution of solute-dislocation interaction energies compared to a purely random sample, including cross-kink separation, which controls the cross-kink strengthening[16]. The balance between these two effects gives rise to the net effects of SRO on strength, which highly depends on the alloy composition, as shown for instance in bcc NbTiZr alloys that exhibit SRO-softening[31]. In the MoNbTaW studied here, the SRO-breaking in individual mechanisms and the resulting energy penalty substantially raise the energy barriers and play a major role in dislocation glide. It is reasonable to speculate that SRO will contribute to the strengthening of this type of alloy at low or cryogenic temperatures. As temperature rises to intermediate range and cross-kinks take over as the dominant role in strengthening of refractory HEAs, the SRO can result in less strengthening because of infrequent kink-pair nucleation (enhanced barrier) and suppressed cross-slip (reduced probability of forming cross-kinks). At high temperatures, thermal diffusion mobilizes cross kinks and the consequent cross-kink annihilation lessens the effectiveness of the strengthening, and therefore, the role of edge dislocations becomes predominant and controls the strength of bcc HEAs[27,32].

Our study reveals and delineates the striking features of rugged energy landscape inherent in refractory HEAs, over which the dislocation motion necessarily operates and proceeds through cooperative elemental mechanisms along the minimum energy pathways. The unveiled PEL demonstrates a multilevel and hierarchical structure, manifesting as a collection of small basins nested in a large metabasin, which is the notable feature making them distinct from their conventional counterparts. By tailoring local chemical distribution and introducing SRO, the energy landscape is appreciably smoothened (reduced variances in local minima and saddle point) yet skewed to different degrees that change the relative role of individual mechanisms as the rate-limiting process for dislocation motion. The energy landscape hierarchy and its high barrier variation, induced by chemical randomness in HEAs, highlight the similarity to that of structural disorder in metallic glasses and render various barrier activation events and dislocation glide mechanisms underpinning the extraordinary behaviors of HEAs.

**Acknowledgments.** X.W. and P.C. acknowledge the start-up grant from the Henry Samueli School of Engineering, University of California, Irvine. The work on SRO formation and analysis is supported by the U.S. Department of Energy (DOE), Office of Basic Energy Sciences, under Award DE-SC0022295.

**Author Contributions.** P.C. conceived the original research idea; X.W. performed research; F.M. and P.C. advised the study; X.W., F.M., and P.C. analyzed the data and wrote the manuscript.

**Declaration of Competing Interests.** The authors declare no competing financial interests.

**Data and code availability**. All relevant data are included within the manuscript and Supplementary Information, and/or are available from the authors. All simulations codes that support the findings of this study are available from Xinyi Wang (xinyiw19@uci.edu) upon reasonable request.

# Figures

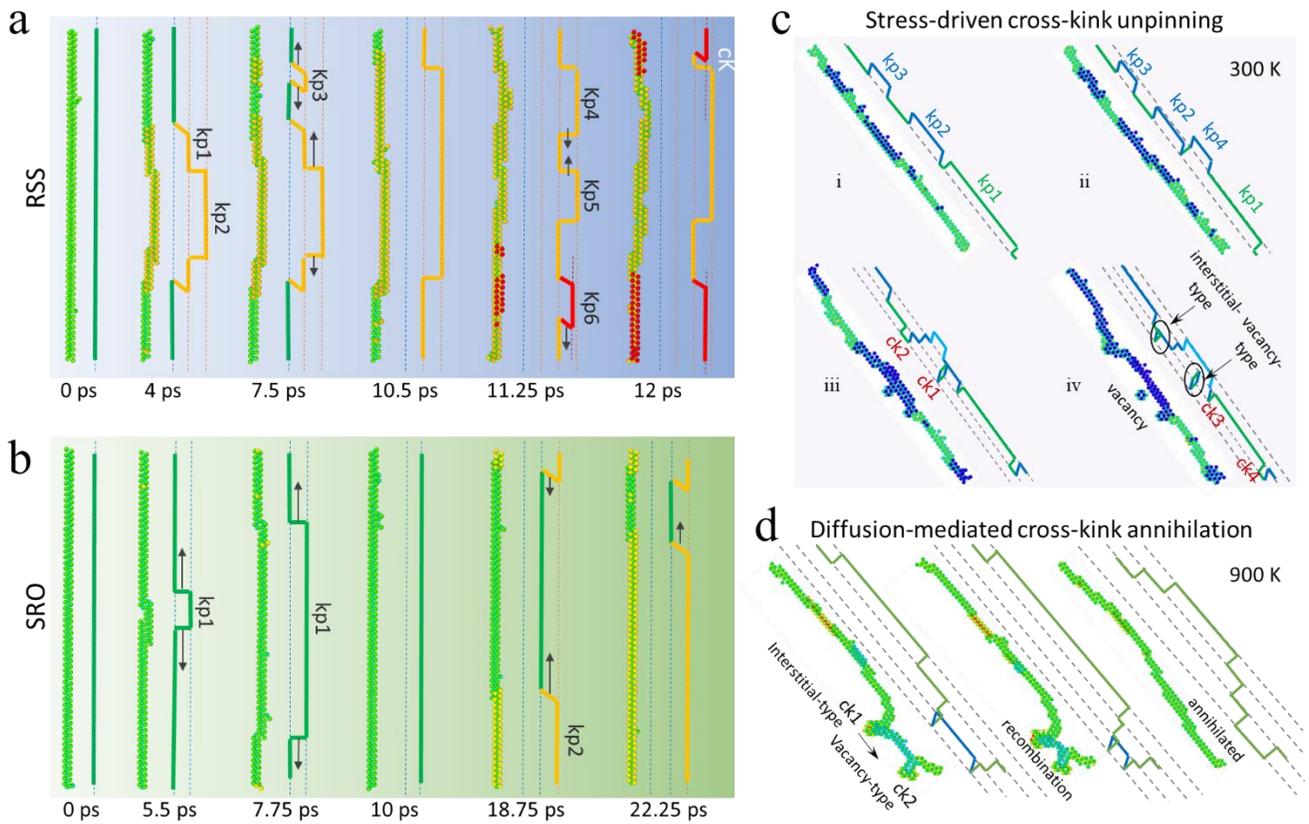

**Figure 1**. Atomic structure evolution and gliding mechanisms of screw dislocation at 300 K and 800 MPa, in (a) RSS and (b) SRO systems. Atoms are colored by their positions along the glide plane normal direction to illustrate cross slip. (c-d) Cross-kink breaking mechanisms at low and intermediate temperatures. (c) stress-driven cross-kink unpinning and production of vacancies and self-interstitials at 300 K, and (d) thermally-activated and diffusion-mediated cross-kink annihilation at 900 K.

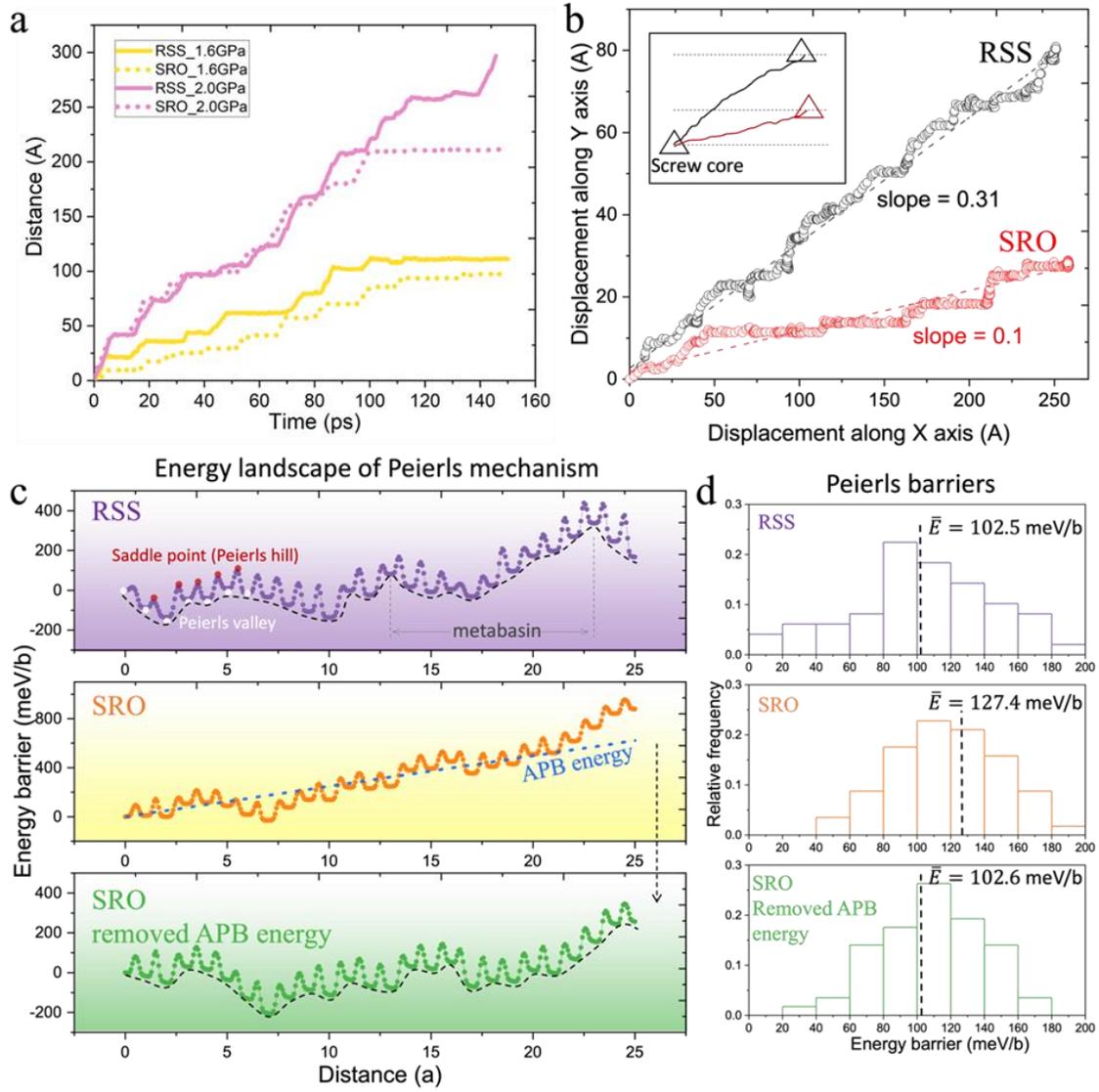

**Figure 2.** Peierls mechanism and its energy landscape. (a) Dislocation glide distance as a function of time for RSS and SRO at two different applied stresses. (b) Trajectory of dislocation core position at 300 K and 2 GPa applied stress. The slope resulting from cross-slip represents the event frequency. (c) Minimum energy pathway and potential energy landscape spanning 25 Peierls valleys in RSS system (purple line). Middle panel shows the energy landscape (orange line) and APB energy due to dislocation glide-induced order breaking (blue). Bottom panel presents the Peierls energy landscape after removing the APB energy. (d) The statistical distributions of energy barrier in RSS, SRO, and SRO with reduced APB energy.

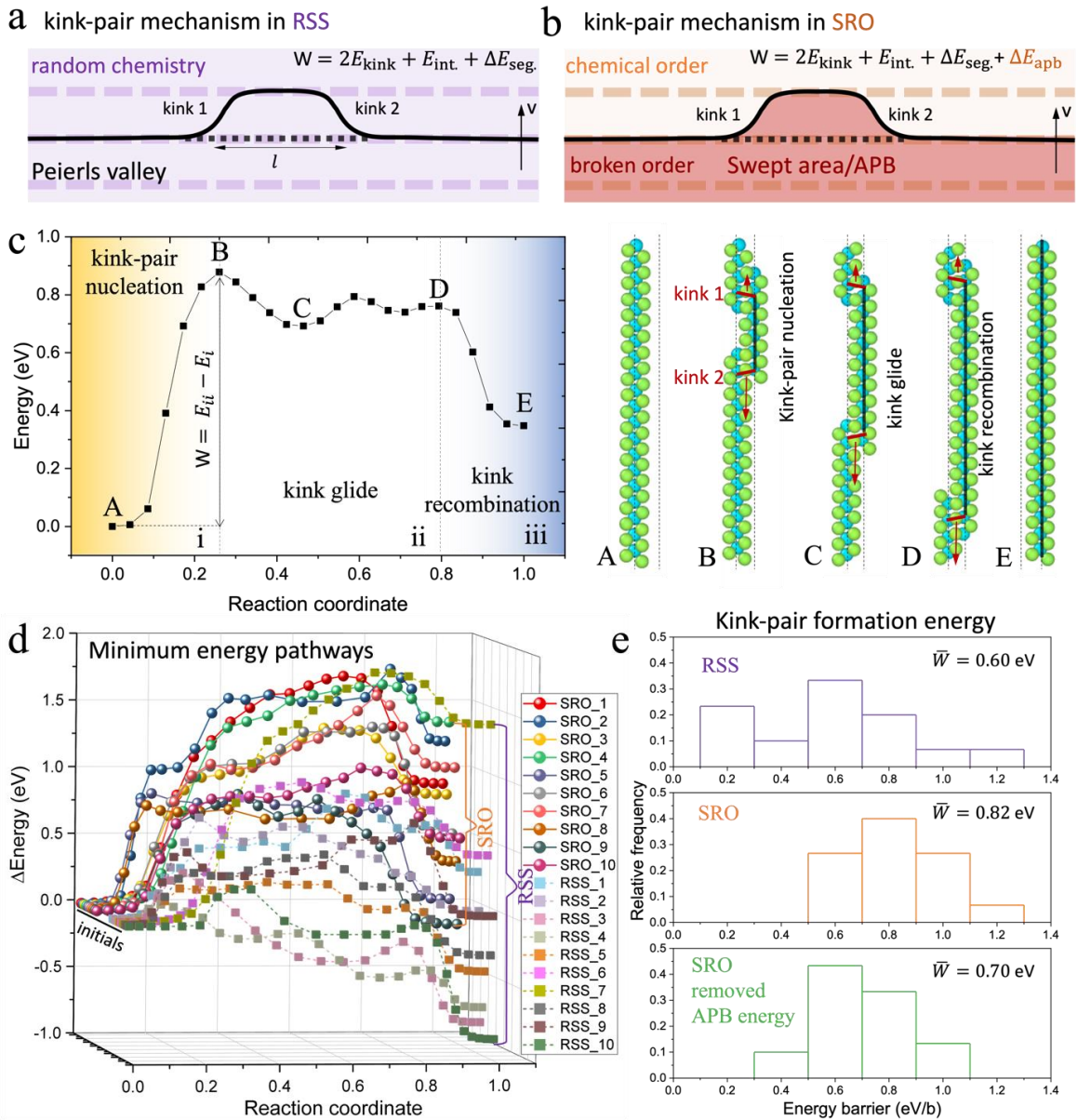

**Figure 3.** Kink-pair nucleation and the energy landscape. (a-b) Schematic illustrations of kink-pair nucleation in RSS and SRO, respectively. (c) Typical minimum energy pathway of kink-pair mechanism and the corresponding structural evolution. A critical-sized double-kink configuration is formed at state B. (d) Twenty minimum energy pathways in RSS and SRO as labeled. (e) The statistical distributions of kink-pair formation energy in RSS, RSO, and SRO with removed APB energy.

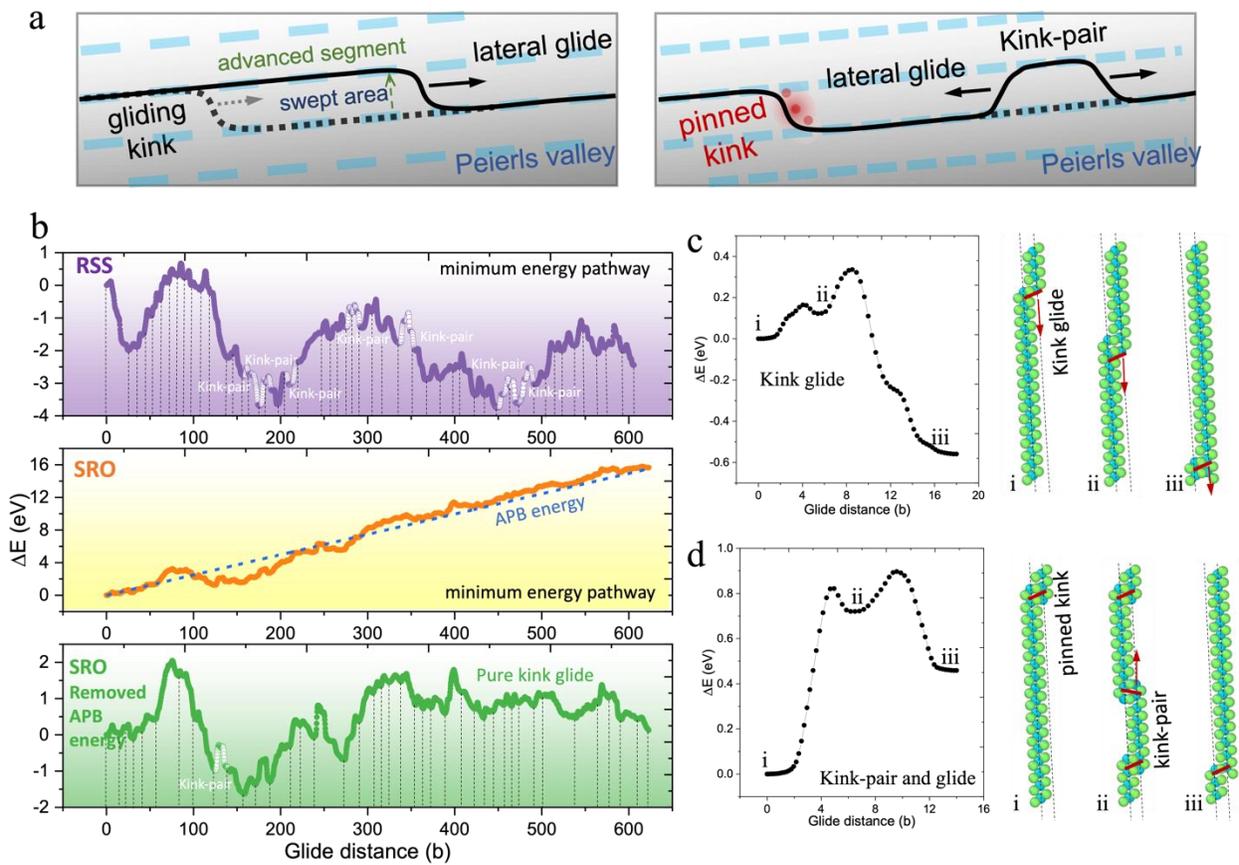

**Figure 4**: Kink glide mechanism and its energy landscape. (a) Schematic illustrations of pure kink glide and pinned kink with kink-pair nucleation. (b) The potential energy landscape of kink gliding over $615b$ distance in RSS (upper panel). The unfilled symbols indicate kink-pair events. Middle panel shows the energy landscape of kink gliding in RSO (orange curve) and the associated APB energy due to order breaking. The green curve is the calculated energy landscape of kink glide in SRO after removing APB energy. (c-d) Typical minimum energy paths and structure evolution of (c) pure kink glide and (d) kink-pair nucleation with chemically pinned kink.
Page 20

# Supplementary information for

The Hierarchical Potential Energy Landscape of Screw Dislocation Motion in Refractory High-entropy Alloys

Xinyi Wang[a,b], Francesco Maresca[c], Penghui Cao[a,b]

[a]Department of Mechanical and Aerospace Engineering, University of California, Irvine, CA 92697, USA

[b]Materials and Manufacturing Technology Program, University of California, Irvine, CA 92697, USA

[c]Engineering and Technology Institute Groningen, Faculty of Science and Engineering, University of Groningen, 9747 AG, Netherlands

This PDF file includes:

    Figures S1 to S11

    Supplementary text

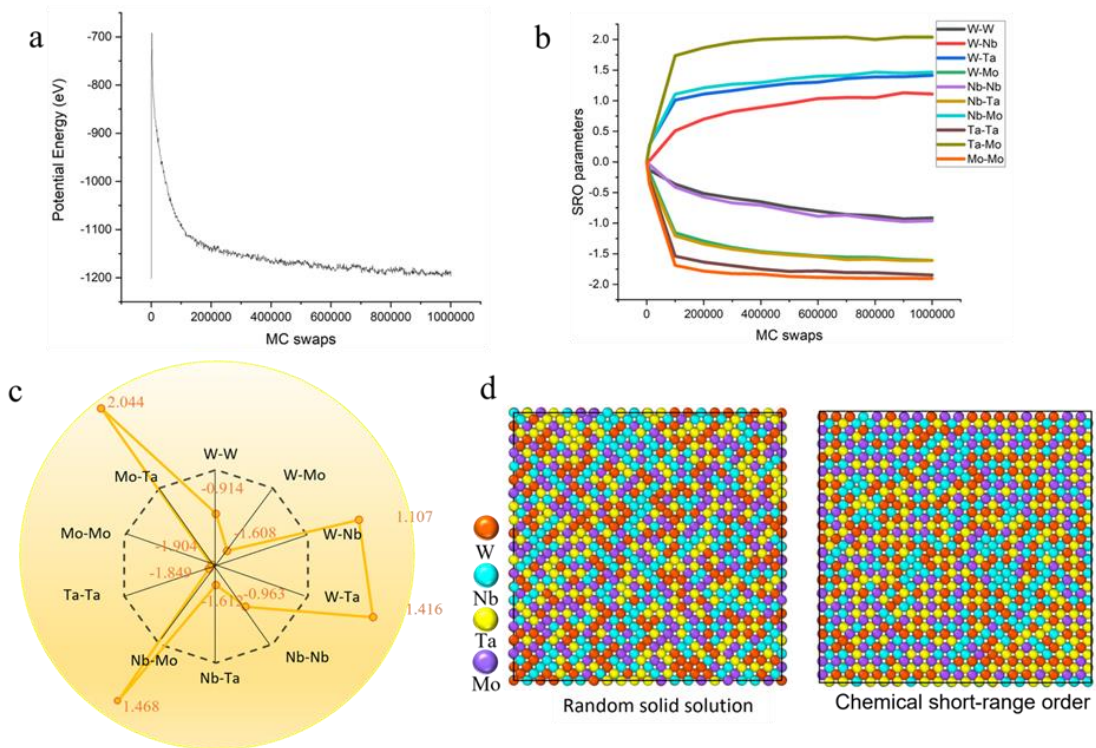

**Figure S1**. MC-MD simulations and chemical SRO parameters. (a) Potential energy change as a function of MC in the hybrid MC-MD simulation. (b) The variation of SRO parameter $\delta_{ij}^{k=1}$ with MC swaps. (c) The calculated SRO parameters at the first nearest neighbor in random solid solution (RSS) and short-range order (SRO) systems. The MC/MD annealed structure exhibits strong ordering between Mo-Ta, Nb-Mo, and W-Ta. (d) The corresponding atomic configurations of RSS and SRO, respectively.

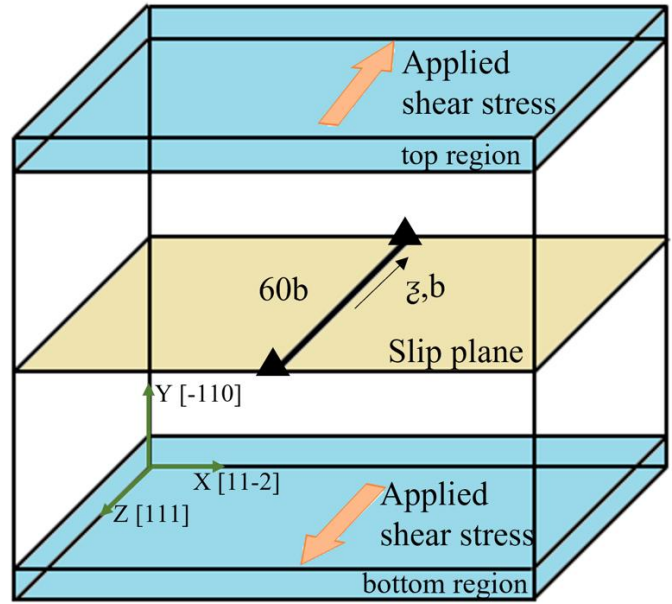

**Figure S2**. MD simulation of screw dislocation model. The crystal has orientations of $[11\bar{2}]$, $[\bar{1}10]$, and $[111]$, along with the $x$, $y$, and $z$ directions. A screw dislocation with Burgers vector $1/2[111]$ is inserted in the middle of the system by Volterra shear and is relaxed using a conjugate gradient method to find its local energy minimum state. The dislocation line direction aligns with $z$ axis, and glide direction along $x$, and the glide plane normal direction $y$. Shear stress is applied to the top and bottom layers of atoms.

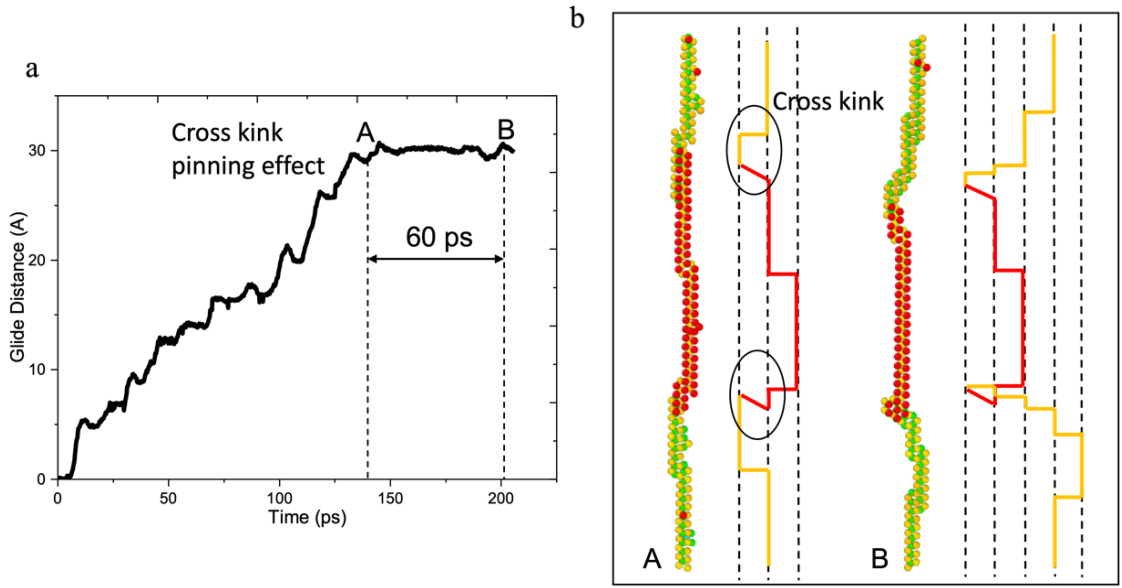

**Figure S3**: Cross kink self-pinning effect. (a) Glide distance versus time for screw dislocation subjected to 0.8 GPa at 300 K in RSS. The plateau (point A to B) is caused by cross kink pinning. (b) Atomic structures of dislocation right before cross-kink formation and pinned state. Two kinks belonging to different planes (yellow and red) collide and generate the cross-kink, leading to self-pinning.

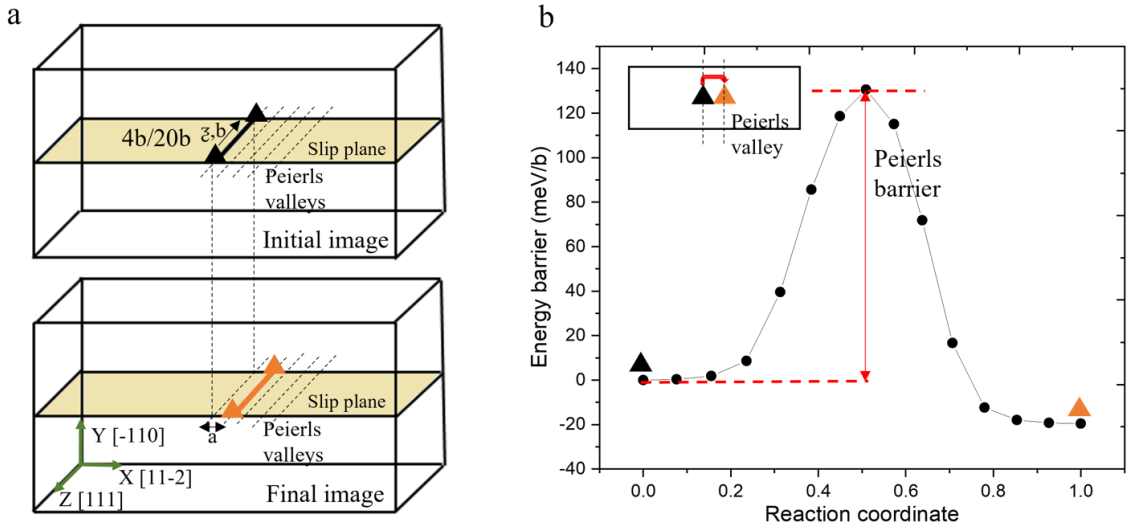

**Figure S4**. Peierls mechanism model and minimum energy path calculation. (a) The initial and final states, which locates at the bottom of Peierls valleys, are used for NEB calculation. (b) The minimum energy pathway and the energy barrier linking two neighboring valleys. For a short dislocation of $4b$, the whole segment hops from the initial valley (black triangle) to the next valley (orange triangle) along the minimum energy pathway.

Note of Figure 4, Peierls mechanism and kink-pair nucleation models. To construct the potential energy landscape, we insert the screw dislocation as the initial state and then move it to the next Peierls valley as the final state. The minimum energy path crossing one Peierls hill is determined by the NEB method (Figure S4b). By performing NEB calculations and connecting the 25 successive valleys, the Peierls energy landscape, including local minima and saddle points, is obtained. Concerning the kink-pair nucleation, the modeled dislocation length is increased to $20b$, which naturally results in one kink-pair formation in the minimum energy path. From that, the kink-pair formation energy is determined (see Fig. 3c in the manuscript).

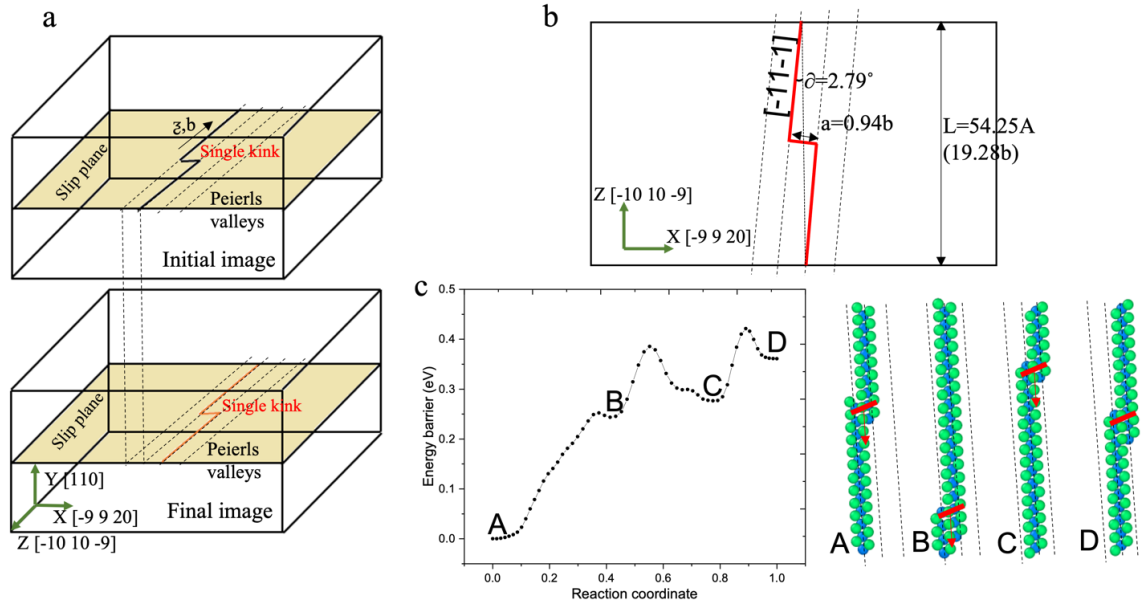

**Figure S5**. Single kink glide model. (a) Simulation setup of kink glide. The periodic boundary conditions are applied along $x$ and $y$ directions. The dashed lines represent Peierls valleys. (b) Relaxation of $[\overline{10}\ 10\ \overline{9}]$ screw dislocation leads to the formation of a single kink that connects two segments located in the Peierls valleys. (c) A typical minimum energy path of kink glide and the corresponding structure evolution of kink propagation.

Note of Fig. S5. For examination of the kink glide energy landscape, we construct a screw dislocation containing a single kink. We first employ a system oriented with x ∥ $[\overline{9}\ 9\ \overline{20}]$, y ∥ $[110]$, and z ∥ $[\overline{10}\ 10\ \overline{9}]$. The angle between $[\overline{10}\ 10\ \overline{9}]$ and $[\overline{1}\ 1\ \overline{1}]$ directions is 2.79°, and the distance between two Peierls valley is $0.94b$ ($b$ is Burgers vector). According to the geometric relationship (L=$0.94b$ /tan∂), the length for forming one single kink is 54.25 Å in the $z$ direction. With this geometry, the initially inserted screw dislocation is relaxed to form a curved structure, as shown in Figure S5b.

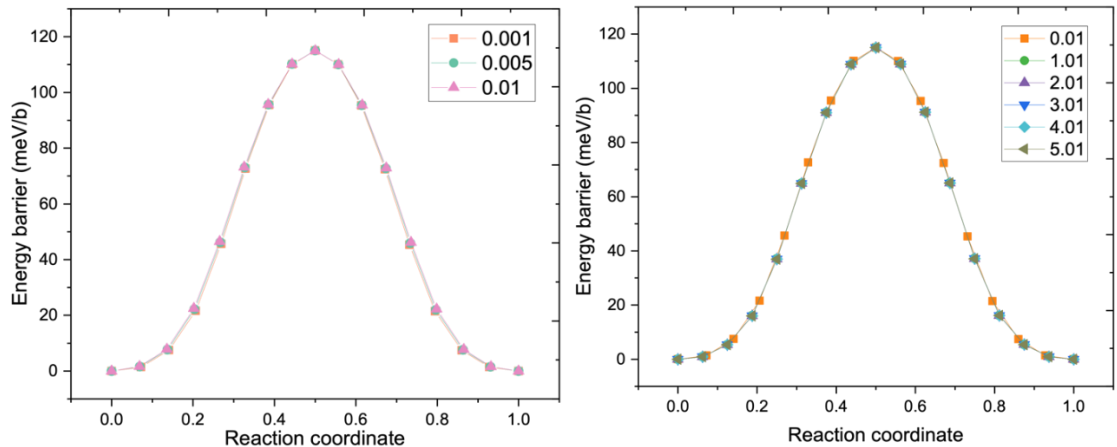

**Figure S6**. NEB parameters testing. (a) The minimum energy paths of Peierls mechanism obtained using various energy tolerances (i.e., 0.001, 0.005, and 0.01 eV/A). (b)The minimum paths with various spring constants, ranging from 0.01 to 5.01 eV/A$^2$.

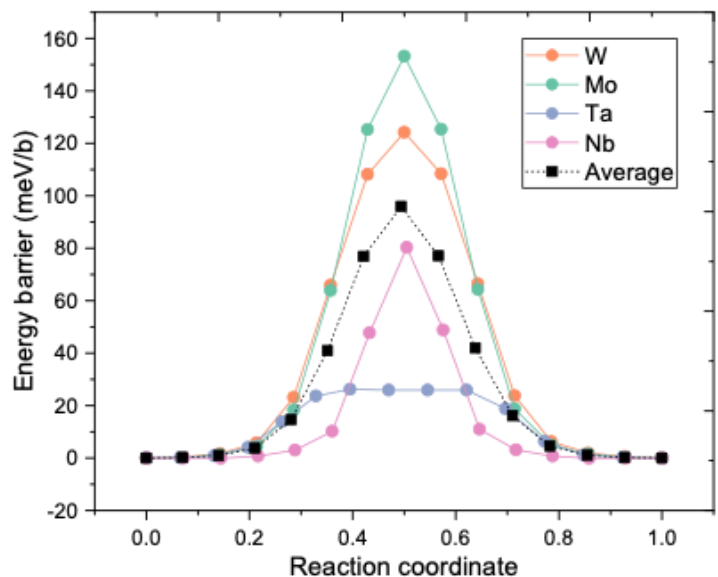

**Figure S7.** Peierls energy barriers of four pure elements (W, Mo, Ta, and Nb). The mean energy barrier of the four elements is about 96 meV/b, which is close to the averaged Peierls barrier in RSS (103 meV/b). The SRO has an overall higher energy barrier (127 meV/b, as shown in Figure 2d), implying an enhanced lattice resistance to dislocation glide in a chemically ordered environment.

# Section 1. Antiphase boundary (APB) energy calculation

Dislocation glide results in chemical order breaking on the slip plane, hence APB generation. To determine the energy cost (APB energy), we shift a slab of atoms (6 layers) in the SRO system by one Burgers vector, $1/2[111]$, as shown in Figure S8. It creates two APBs, indicated by red dashed lines. The configuration with APBs is then relaxed with energy minimization to obtain the potential energy. The energy of APB is calculated as

$$E_{APB} = (PE_{APB} - PE_0)/2$$

where $PE_0$ is the system energy before slip and $PE_{APB}$ is the potential energy after slip. By normalizing the slip area, the APB energy per area is determined. We repeat the slip at five different locations, and the averaged APBs energy per area is calculated to be 53.5 mJ/m$^2$.

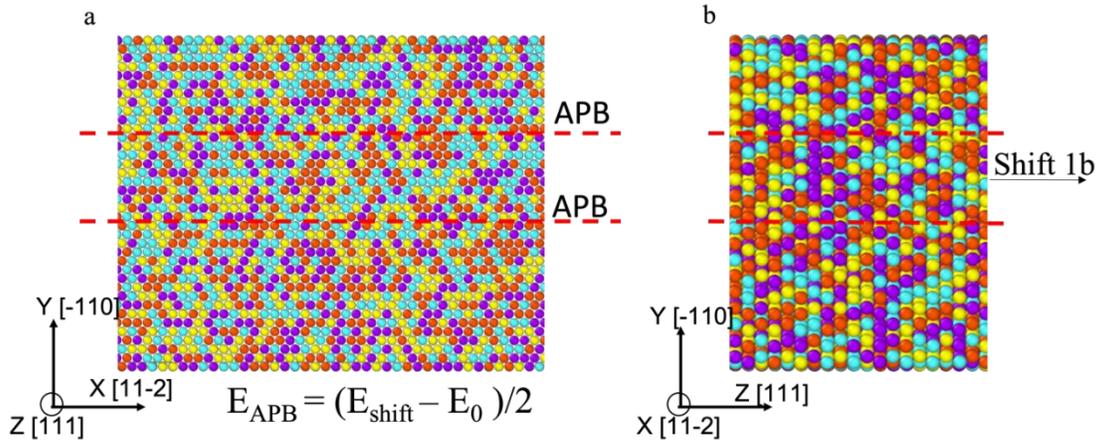

**Figure S8.** Calculation APB energy. (a) Perfect crystal with two red dashed lines representing two generated antiphase boundaries. (b) Side view of the same system, no defects created after the shift.

## Section 2. Screw dislocation core energy

The screw dislocation core energy for RSS can be calculated by the following equation,

$$E_{core}^{RSS} = (E_{screw} - E_0)/4b$$

where $E_0$ is the energy of system without screw dislocation, $E_{screw}$ is the energy of the system containing one screw dislocation, and $4b$ is the dislocation length. In the presence of SRO, the creating of screw dislocation causes slip (i.e., APB generation). The APB energy should be excluded when considering dislocation core energy,

$$E_{core}^{SRO} = (E_{screw} - E_0 - E_{APB})/4b$$

where $E_{APB}$ is the corresponding APB energy behind the dislocation in the SRO environment. Figure S9 shows the statistical distributions of screw dislocation core in RSS and SRO, respectively. The presence of SRO does not change the average values of core energies but tends to narrow the core energy distribution.

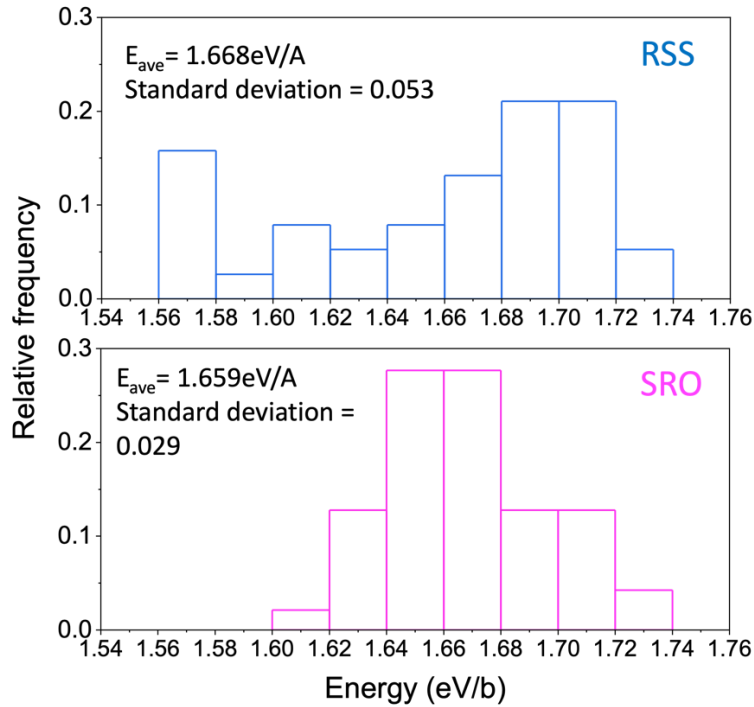

**Figure S9**. Statistical distribution of screw dislocation core energy in RSS and SRO systems**.** The presence of SRO narrows down the distribution of the dislocation core energy (i.e., reducing the variance of Peierls valley levels from the PEL perspective).

**Section 3. Local energy minima and saddle point detection**

We use SciPy Python package to find all local minima and saddle points on the minimum energy pathway. The prominence of peaks measures how much a peak stands out from the surrounding baseline and is calculated by the vertical distance between local peaks and their neighboring local minima. Here we use 0.03 eV as the threshold value for prominence, and the small peaks are regarded as noise (Figure S10). We also test different threshold values (0.08 eV and 0.1 eV). The larger values can smear out less prominent peaks, but the overall trend of results remains the same. Figure S8 shows the potential energy landscape of kink glide in RSS with all marked local minima and saddle points. After extracting these values, the local energy barriers and distances connecting two local minima can be determined, as displayed in Figure S11. The kink glide in SRO experiences an overall larger energy barrier ($E_{SRO} = 0.61\ eV$), as compared to RSS ($E_{RSS} = 0.43\ eV$). Figure S11b shows distributions of the distance between adjacent local minima. RSS exhibits a shorter average distance between two local minima than that of SRO, which can be attributed to the large chemical fluctuations and strong solute-kink interaction in RSS.

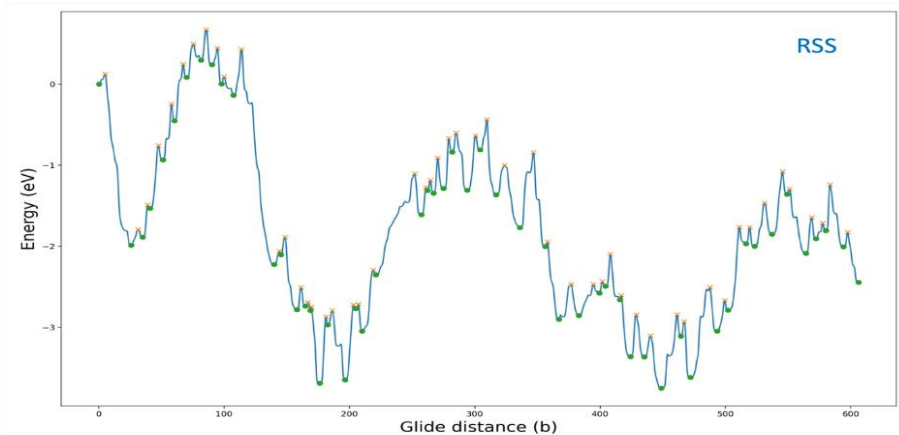

**Figure S10**. The detected local minima and saddle points in the minim energy path of kink glide. These critical points are extracted using a threshold value of 0.03 eV. The length scale (i.e., distance separation between local minima) and energy barrier can be obtained.

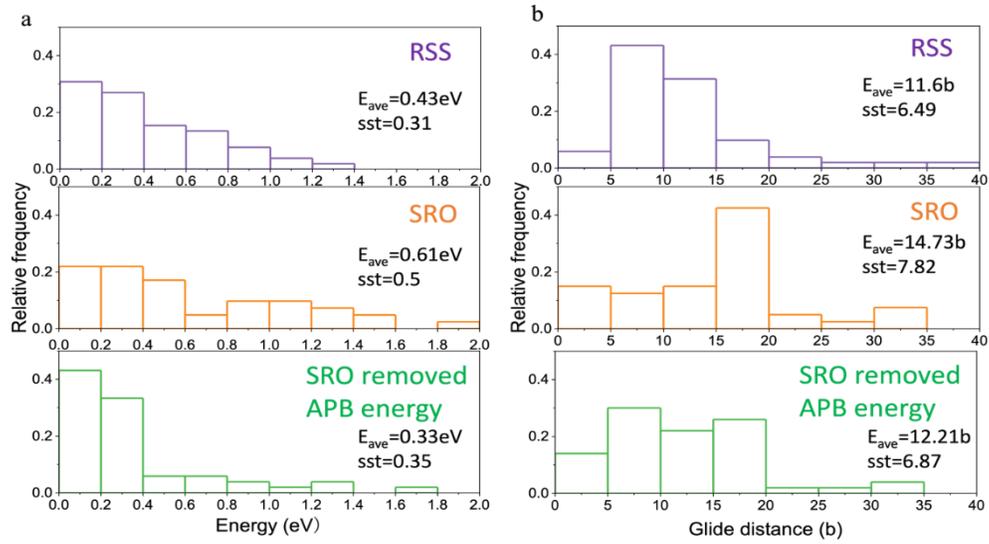

**Figure S11**. The energy barrier and length scale of kink glide in RSS and SRO. (a) Energy barrier distributions of kink glide in RSS, SRO, SRO-APB, respectively. (b) Distributions of local minimum distance in RSS, SRO, and SRO-APB, respectively. The SRO shows a longer length scale (periodicity) between local energy minima of kinked configuration compared to RSS.